\documentclass[a4paper,hyper,cits]{PoS-mod}
\usepackage{url,amsmath,amssymb,multirow}

\title{$R$-parity violating supersymmetry and neutrino physics: experimental signatures}

\ShortTitle{$R$-parity violating supersymmetry and neutrino physics}

\author{\speaker{Vasiliki A.\ Mitsou}\\
        Instituto de F\'isica Corpuscular (IFIC), CSIC -- Universitat de Val\`encia, \\ 
Parc Cient\'ific de la U.V., C/ Catedr\'atico Jos\'e Beltr\'an 2, \\
E-46980 Paterna (Valencia), Spain\\
and\\
CERN, PH Department, CH-1211 Geneva 23, Switzerland\\
        E-mail: \email{vasiliki.mitsou@ific.uv.es}}

\abstract{$R$-parity violating supersymmetric models (RPV SUSY) are becoming increasingly more appealing than its $R$-parity conserving counterpart in view of the hitherto non-observation of SUSY signals at the LHC. In this paper, we discuss RPV scenarios where neutrino masses are naturally generated, namely RPV through bilinear terms (bRPV) and the $\mu$-from-$\nu$ supersymmetric standard model ($\mu\nu$SSM). The latter is characterised by a rich Higgs sector that easily accommodates a 125-GeV Higgs boson. The phenomenology of such models at the LHC is reviewed, giving emphasis on final states with displaced objects, and relevant results obtained by LHC experiments are presented. The implications for dark matter for these theoretical proposals is also addressed.}

\FullConference{18th International Conference From the Planck Scale to the Electroweak Scale \\
		25-29 May 2015\\
		Ioannina, Greece }
\def\tev{\textrm{TeV}}
\def\gev{\textrm{GeV}}
\def\pt{\ensuremath{p_{\rm T}}}
\def\met{\ensuremath{E_{\rm T}^{\rm miss}}}

\def\ifb{\ensuremath{\textrm{fb}^{-1}}}
\def\to{\ensuremath{\rightarrow}}
\def\X{\ensuremath{\tilde\chi_1^0}}

\def\t#1{\tilde{ #1}}

\def\mn{\ensuremath{\mu\nu}SSM}
\def\XX{\ensuremath{\tilde\chi_4^0}}
\DeclareGraphicsExtensions{.pdf}
\begin{document}

%
%%%%%%%%%%%%%%%%%%%%%%%%%%%%%%%%%%%%%%%%%%%%%%%%%%%%%%%%%%%%%%%%%%%%%%%%%%%%%%%%%%%%%%%%%%%%%%%%%%%%
%%%%%%%%%%%%%%%%%%%%%%%%%%%%%%%%%%%%%%%%%%%%%%%%%%%%%%%%%%%%%%%%%%%%%%%%%%%%%%%%%%%%%%%%%%%%%%%%%%%%
\section{Introduction}\label{sc:intro}

Supersymmetry (SUSY)~\cite{susy} is an extension of the Standard Model (SM) that assigns to each SM field a superpartner field with a spin differing by half a unit. SUSY provides elegant solutions to several open issues of the SM, such as the hierarchy problem, the nature of dark matter~\cite{dm-review}, and the grand unification. SUSY is one of the most relevant scenarios of new physics probed at the LHC, yet no signs of SUSY have been observed so far. In view of these null results in \emph{conventional} SUSY searches, it becomes  mandatory to fully explore \emph{non-standard} SUSY scenarios involving $R$-parity violation (RPV)~\cite{rpv} and/or quasi-stable particles. 

$R$~parity is defined as $R = (-1)^{3(B-L)+2S}$, where $B$ ($L$) is the baryon (lepton) number and $S$ the spin, respectively, granting $R=+1$ ($R=-1$) to all SM particles (SUSY partners). It is worth emphasising that the conservation of $R$~parity is merely an \emph{ad-hoc} assumption with the only strict limitation coming from the proton lifetime: non-conservation of both $B$ and $L$ leads to a rapid proton decay. $R$-parity conservation has serious consequences in SUSY phenomenology in colliders: the SUSY particles are produced in pairs and, most importantly, the lightest supersymmetric particle (LSP) is absolutely stable and weakly interacting, thus providing the characteristic high transverse missing momentum (\met) in SUSY events at colliders. Here we highlight two RPV scenarios: the bilinear RPV (bRPV) and the $\mu$-from-$\nu$ supersymmetric standard model (\mn), which both reproduce correctly the neutrino physics observations.

The paper is structured as follows. Section~\ref{sc:brpv} is dedicated to bRPV SUSY models, while the $\mn$ is discussed in Section~\ref{sc:munussm}. Aspects of RPV SUSY linked to dark matter are hightlighted in Section~\ref{sc:dm}. The paper concludes with a summary and an outlook in Section~\ref{sc:summary}.

%%%%%%%%%%%%%%%%%%%%%%%%%%%%%%%%%%%%%%%%%%%%%%%%%%%%%%%%%%%%%%%%%%%%%%%%%%%%%%%%%%%%%%%%%%%%%%%%%%%%
%%%%%%%%%%%%%%%%%%%%%%%%%%%%%%%%%%%%%%%%%%%%%%%%%%%%%%%%%%%%%%%%%%%%%%%%%%%%%%%%%%%%%%%%%%%%%%%%%%%%
\section{Bilinear $R$-parity breaking}\label{sc:brpv}

$R$-parity conservation (RPC) has several consequences such as the stability of the LSP, which is a weakly interacting massive particle (WIMP) and consequently a candidate for dark matter (DM)~\cite{dm-review}. Being a WIMP, once produced at the LHC, it will escape detection, resulting in large missing transverse momentum, \met. Providing a DM candidate is one of the strongest arguments in favour of RPC SUSY, nonetheless RPV models \emph{do} exist that can explain DM through, for instance, very light gravitinos~\cite{gravitino,martin,brpv-dm,brpv-split,ams02-brpv,grefe,steffen,trpv-gravitino}, axions~\cite{axion,steffen} or axinos~\cite{steffen,axino,brpv-axino}. 

As long as the breaking of $R$~parity is spontaneous, only bilinear terms arise in the effective theory below the RPV scale, thus rendering bilinear $R$-parity violation (bRPV) a theoretically attractive scenario. Moreover, the bilinear model provides a self-consistent scheme in the sense that trilinear RPV implies that also bilinear RPV is present, but not conversely~\cite{Porod:2000hv}.  In other words, the simplest way to break $R$~parity is to add bilinear terms to the MSSM potential. Besides that, additional bilinear soft SUSY breaking terms are introduced, which include small vacuum expectation values for the sneutrinos. In fact, if SUSY was not broken, the bilinear terms could be rotated away and be converted into trilinear terms, however the presence of soft SUSY breaking terms gives bRPV a physical meaning~\cite{Romao:1999up}. 

%%%%%%%%%%%%%%%%%%%%%%%%%%%%%%%%%%%%%%%%%%%%%%%%%%%%%%%%%%%%%%%%%%%%%%%%%%%%%%%%%%%%%%%%%%%%%%%%%%%%
\subsection{Connection with neutrino physics}\label{sc:brpv-nu}

Sneutrino VEVs introduce a mixing between neutrinos and neutralinos, leading to a see-saw mechanism that gives mass to one neutrino mass scale at tree level with the second neutrino mass scale being induced by loop effects~\cite{Hirsch:2000ef,Diaz:2003as}. The same VEVs are also involved in the decay of the LSP, which in this case is the lightest neutralino. This implies a relation between neutrino physics and some features of the LSP modes. An example of such a connection is given by the relation~\cite{Hirsch:2000ef}
\begin{equation}
\label{tetatm}
\tan^2\theta_{\rm atm} = \left|{\frac{\Lambda_{\mu}}{\Lambda_{\tau}}}\right|^2 
\simeq \frac{BR(\tilde\chi_1^0\to\mu^\pm W^\mp)}{BR(\tilde\chi_1^0\to\tau^\pm W^\mp)} 
= \frac{BR(\tilde\chi_1^0\to\mu^\pm q \bar{q}')}{BR(\tilde\chi_1^0\to\tau^\pm q \bar q')},
\end{equation}
 
\noindent where $\theta_{\rm atm}$ is the atmospheric neutrino mixing angle and the ``alignment'' parameters $\Lambda_i$ are defined as $\Lambda_i = \mu v_i + v_d \epsilon_i$, with $v_d$ the VEV of $H^d$. This relation between RPV and neutrino physics allows setting bounds on bRPV parameters from results of neutrino experiment and astrophysical data~\cite{Abada:2000xr}. In the opposite direction, a possible positive signal observed in colliders may lead to the determination of some of the bRPV phenomenological properties, which in turn can constrain neutrino-physics parameters~\cite{brpv-nu}.

%%%%%%%%%%%%%%%%%%%%%%%%%%%%%%%%%%%%%%%%%%%%%%%%%%%%%%%%%%%%%%%%%%%%%%%%%%%%%%%%%%%%%%%%%%%%%%%%%%%%
\subsection{Phenomenology at the LHC}\label{sc:brpv-pheno}

In the bilinear $R$-parity violating models discussed here, the LSP is the lightest neutralino, $\tilde\chi_1^0$~\cite{deCampos:2007bn}. The six bRPV parameters for each chosen model point are determined by the \texttt{SPheno}~\cite{spheno} spectrum calculator. \texttt{SPheno} uses as input the model parameters and the neutrino physics constraints, and delivers as output the bRPV parameters (together with the mass spectra and the decay modes) compatible with these constraints. Once those quantities are calculated for a given set of bRPV parameters within a model point, they can subsequently passed to an event generator and produce collision events at the LHC.  

\begin{figure}[htb]
\begin{minipage}{0.3\textwidth}
\includegraphics[width=\textwidth]{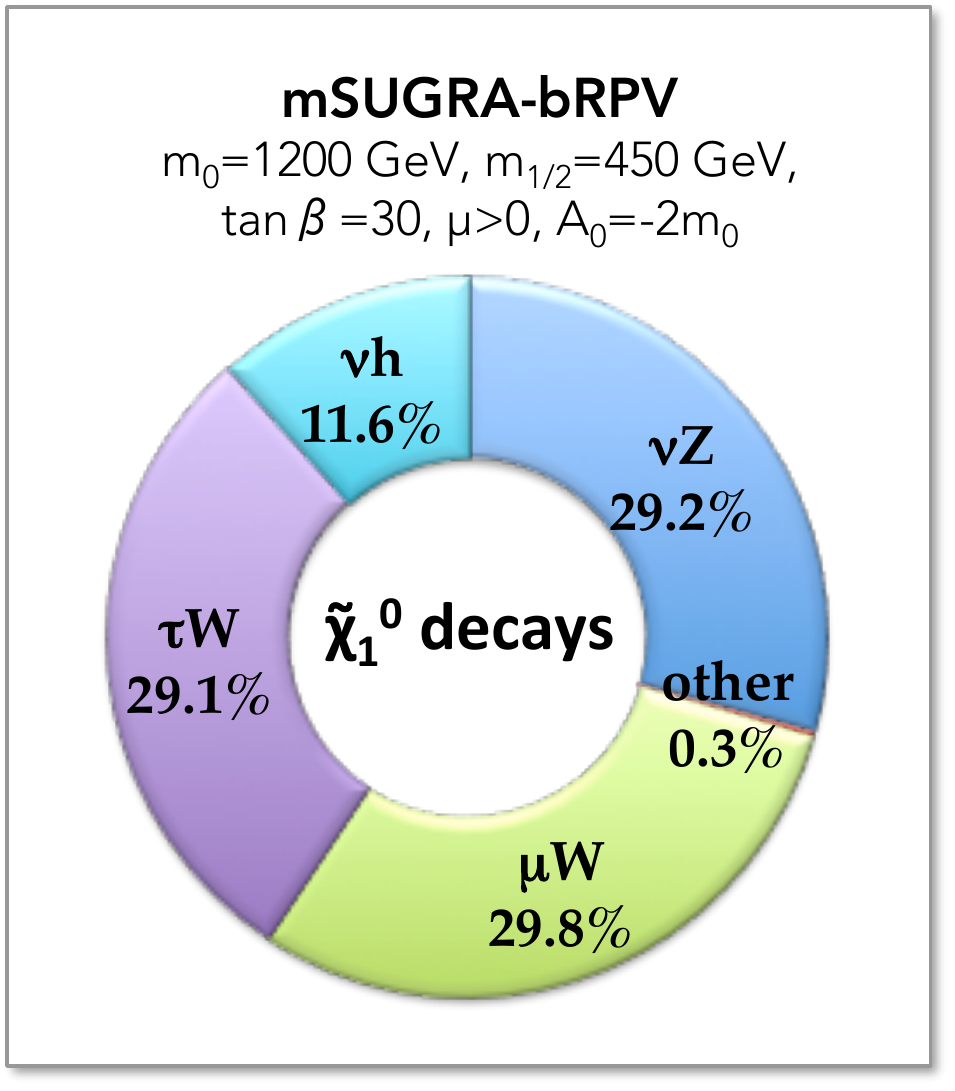}
\caption{\label{fg:decay}\X\ decay modes for a bRPV-mSUGRA point.}
\end{minipage}
\hfill
\begin{minipage}{0.63\textwidth}
\centering
\includegraphics[width=0.8\textwidth]{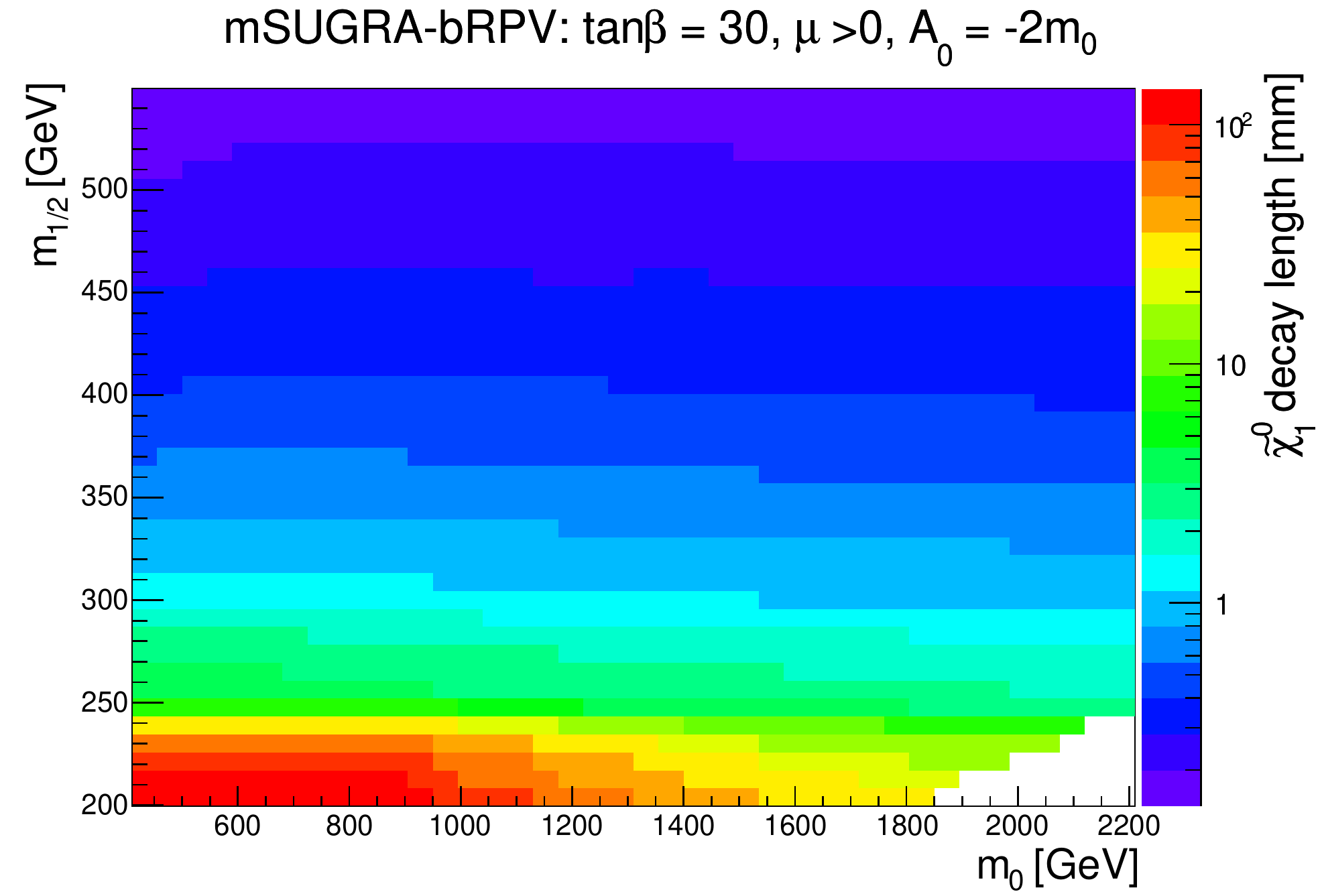}
\caption{\label{fg:decay-length}Proper decay length $c\tau$ in millimetres for the lightest neutralino LSP in bRPV-mSUGRA parameter plane $(m_0,\,m_{1/2})$.}
\end{minipage} 
\end{figure}

To illustrate the basic features of the bRPV phenomenology at LHC, we embed the bRPV terms to the minimal Supergravity (mSUGRA) model. Similarly such terms have been introduced to anomaly-mediated SUSY breaking~\cite{brpv-amsb} and the natural pMSSM model, discussed in Section~\ref{sc:brpv-atlas}. The sparticle spectrum for bRPV-mSUGRA is ---within theoretical uncertainties--- the same as in RPC mSUGRA; it is the LSP decay modes and its lifetime that depend on the bRPV parameters. Typical \X\ decay modes are shown in Fig.~\ref{fg:decay} for a model compatible with a \mbox{125-\gev} Higgs boson~\cite{atlas-higgs,cms-higgs}. The neutralino decays are dominated by leptonic $(e,\,\mu)$ and $\tau$ channels, making lepton-based searches ideal for constraining this model~\cite{int-note,emma,brpv-nu}. Furthermore the fact that in the low-$m_{1/2}$ high-$m_0$ region the \X\ is slightly long lived, as evident from Fig.~\ref{fg:decay-length}, opens up the possibility to use searches for displaced vertices~\cite{atlas-dv,atlas-dv-dilep} in order to probe this model at the LHC~\cite{brpv-dv}. Lastly, the \X\ decays to one or two neutrinos give rise to a moderately high \met, thus rendering some \met-based analyses pertinent to bRPV.

%%%%%%%%%%%%%%%%%%%%%%%%%%%%%%%%%%%%%%%%%%%%%%%%%%%%%%%%%%%%%%%%%%%%%%%%%%%%%%%%%%%%%%%%%%%%%%%%%%%%
\subsection{Constraints from ATLAS}\label{sc:brpv-atlas}

Apart from relatively high \met, high lepton/$\tau$ multiplicity is also expected from the LSP decays and from upstream lepton production in the SUSY cascade decay, if strong production is considered. Both features are exploited when looking for a signal of the bRPV-mSUGRA model~\cite{int-note,brpv-atlas1,brpv-atlas2,vam}. The very first bounds set on a bilinear RPV model in colliders were provided by an inclusive search for high \met, jets and one muon at $\sqrt{s}=7~\tev$ and $\sim 1~\ifb$ of ATLAS data~\cite{brpv-atlas1,emma}, which were further extended with $5~\ifb$ by an analysis based on events with high jet multiplicity, large \met\ and exactly one lepton~\cite{brpv-atlas2,elena}.

Further constraints on bRPV-SUGRA for parameters tuned to attain a mass of the lightest Higgs equal to 125~\gev\ have been set recently by ATLAS~\cite{atlas-ss-lep,atlas-tau,atlas-leptons,atlas-strong} with the full data set of $\sim 20~\ifb$ recorded at $\sqrt{s}=8~\tev$. Analyses benefiting from LSP decays to taus~\cite{atlas-tau} and the production of two same-sign leptons or three $b$-jets~\cite{atlas-ss-lep} have been combined~\cite{atlas-strong} to acquire the most up-to-date bounds on bRPV-mSUGRA, shown in Fig.~\ref{fg:atlas-strong}. Similar limits have been set by ATLAS by requiring one hard lepton, jets and large \met~\cite{atlas-leptons}. 
                  
\begin{figure}[htb]
\centering
\begin{minipage}{0.475\textwidth}
\includegraphics[width=\textwidth]{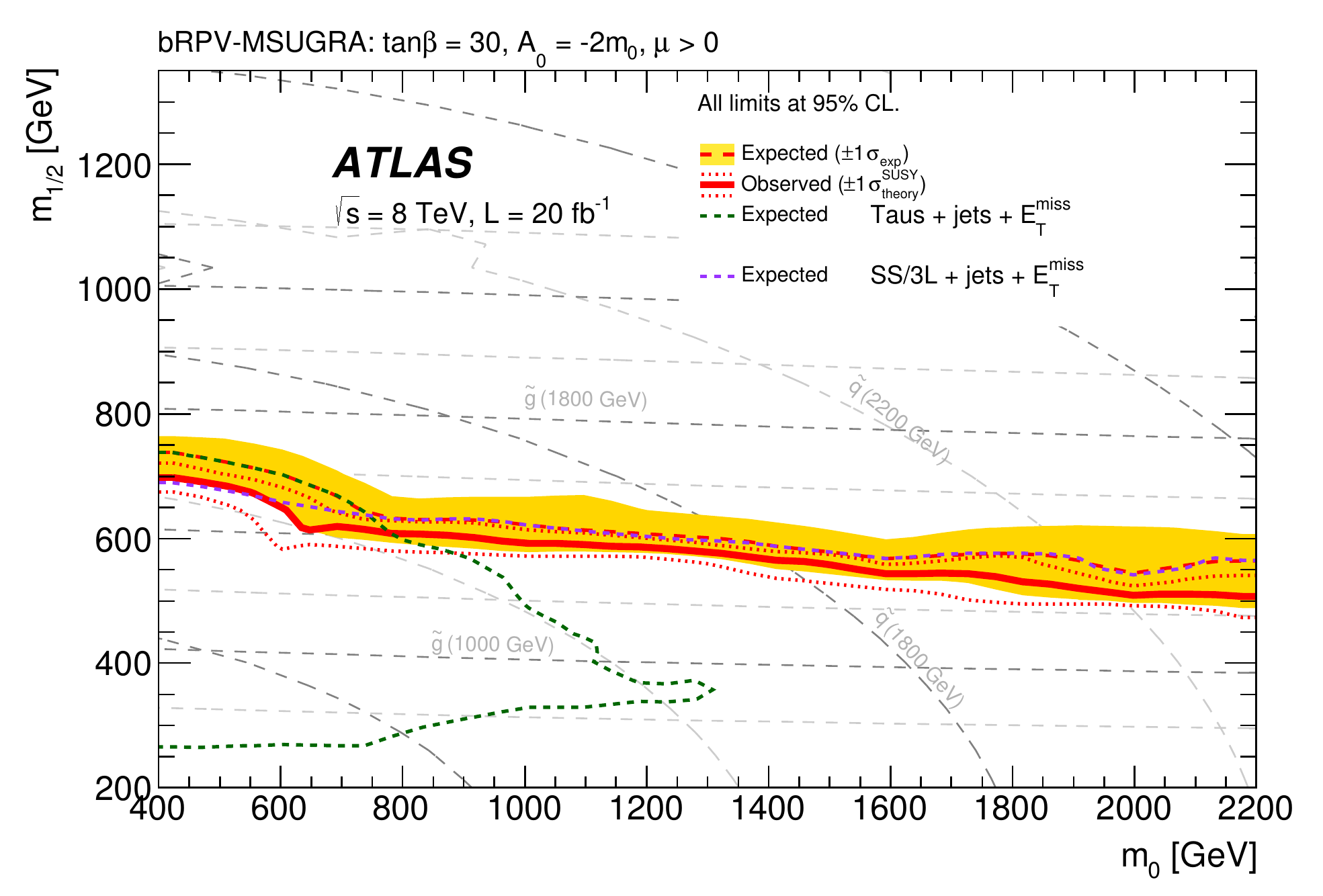}
\caption{Exclusion limits in the $(m_0,\,m_{1/2})$ plane for the bRPV-mSUGRA model. The solid red line and the dashed red line show respectively the combined observed and expected 95\% CL exclusion limits. From Ref.~\cite{atlas-strong}. } \label{fg:atlas-strong} 
\end{minipage}
\hfill
\begin{minipage}{0.475\textwidth}
\includegraphics[width=\textwidth]{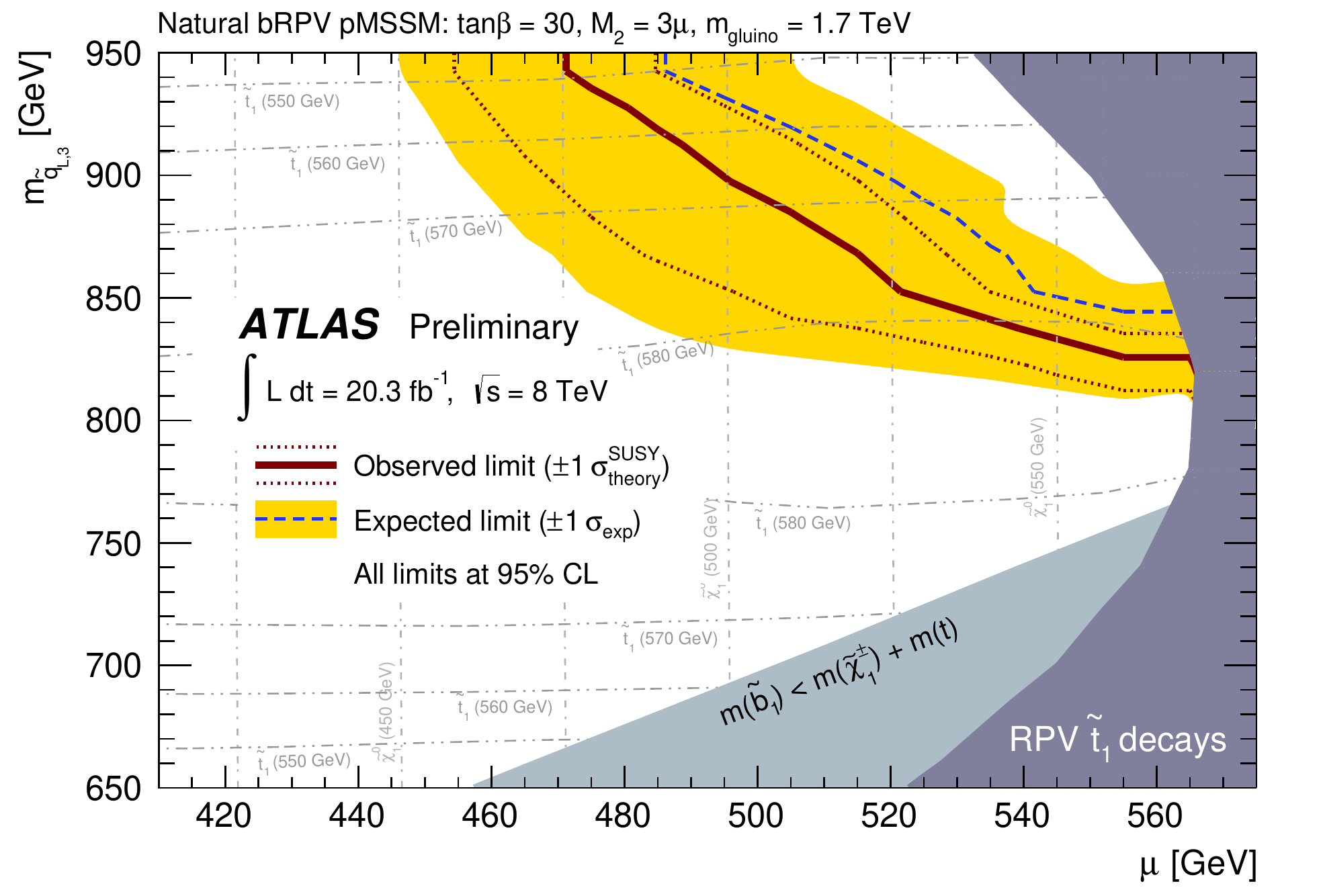}
\caption{Observed (solid) and expected (dashed) 95\% CL exclusion limits for the natural-pMSSM bRPV model, obtained by the SS/3L analysis~\cite{atlas-ss-lep}. The shaded areas at high $\mu$ are not explored here. From Ref.~\cite{atlas-rpv-summary}. } \label{fg:atlas-pmssm} 
\end{minipage} 
\end{figure}

The fact that in a large part of the parameter space the \X\ LSP decays to a $Z$~boson and a neutrino implies that searches for events containing $Z$ and large \met~\cite{cms-z,atlas-z} should be pertinent to explore bRPV as well. In fact this LSP decay ---as much as decays to $\tau W$ and $\mu W$--- is considerable not only when bRPV is embedded in mSUGRA, but also when bRPV is introduced to other well-motivated SUSY scenarios such as ``natural'' SUSY~\cite{nat-susy} and pMSSM~\cite{pmssm}. An interpretation of the results of the same-sign lepton analysis~\cite{atlas-ss-lep} in terms of natural pMSSM~\cite{nat-pmssm} with bRPV has led to the first ever bounds on bRPV within pMSSM, shown in Fig.~\ref{fg:atlas-pmssm}~\cite{atlas-rpv-summary}.

%%%%%%%%%%%%%%%%%%%%%%%%%%%%%%%%%%%%%%%%%%%%%%%%%%%%%%%%%%%%%%%%%%%%%%%%%%%%%%%%%%%%%%%%%%%%%%%%%%%%
%%%%%%%%%%%%%%%%%%%%%%%%%%%%%%%%%%%%%%%%%%%%%%%%%%%%%%%%%%%%%%%%%%%%%%%%%%%%%%%%%%%%%%%%%%%%%%%%%%%%
\section{The $\mu$-from-$\nu$ supersymmetric standard model ($\mu\nu$SSM)}\label{sc:munussm}

The \mn~\cite{mnssm-proposal,mnssm-spectrum} is a theoretical proposal that solves the $\mu$~problem~\cite{mu-problem} of the minimal supersymmetric standard model (MSSM) using the $R$-parity breaking couplings between the right-handed neutrino superfields and the Higgs bosons in the superpotential, $\lambda_i\hat{\nu}_i^c\hat{H}_d\hat{H}_u$. The $\mu$~term is generated spontaneously through sneutrino vacuum expectation values, $\mu = \lambda_i\langle\t{\nu}_i^c\rangle$, once the electroweak symmetry is broken, without introducing an extra singlet superfield as in the case of the next-to-MSSM (NMSSM)~\cite{nmssm}. The complete \mn\ superpotential is given by
%
%%%%%%%%%%%%%%%%%%%%%%%%%%%%%%%%%%%%%%%%%%
%{\small
\begin{equation}\label{eq:w}
\begin{aligned}
W  = &
\ \epsilon_{ab} (
Y_{u_{ij}} \, \hat H_u^b\, \hat Q^a_i \, \hat u_j^c +
Y_{d_{ij}} \, \hat H_d^a\, \hat Q^b_i \, \hat d_j^c +
Y_{e_{ij}} \, \hat H_d^a\, \hat L^b_i \, \hat e_j^c 
\\ 
&
+ Y_{\nu_{ij}} \, \hat H_u^b\, \hat L^a_i \, \hat \nu^c_j -   
\lambda_{i} \, \hat \nu^c_i\,\hat H_d^a \hat H_u^b)+
\frac{1}{3}
\kappa{_{ijk}} 
\hat \nu^c_i\hat \nu^c_j\hat \nu^c_k.
\end{aligned}
\end{equation}
%}
%%%%%%%%%%%%%%%%%%%%%%%%%%%%%%%%%%%%%%%%%%%%%

The couplings $\kappa_{ijk}\hat{\nu}_i^c\hat{\nu}_j^c\hat{\nu}_k^c$ forbid a global $U(1)$ symmetry avoiding the existence of a Goldstone boson, and also contribute to spontaneously generated Majorana masses for neutrinos at the electroweak scale. The latter feature is unlike the bilinear RPV model, where, as mentioned in Section~\ref{sc:brpv-nu}, only one mass is generated at the tree level and loop corrections are necessary to generate at least a second mass and a neutrino mixing matrix compatible with experiments. Moreover in bRPV, the $\mu$-like problem~\cite{mu-problem-nilles} is augmented by the presence of three bilinear terms.

The \mn\ phenomenology is largely defined by the parameters
\begin{equation}
\pmb{\lambda},\:\kappa_{i}, \:\nu^c,\: \tan\beta, \:M_1, \:A_{\lambda}, \: A_\kappa\, ,
\label{EWF-param3A}
\end{equation}
where $\pmb{\lambda}\equiv\sqrt{3}\lambda$ it the singlet-doublet mixing parameter (if universal $\lambda_i$ are assumed), $\kappa$ is the  common $\kappa_{ijk}$, $A_{\lambda}, A_{\kappa}$ are the soft SUSY-breaking parameters and $M_1$ is the $U(1)$ mass scale.

In the \mn, as a consequence of $R$-parity violation, all neutral fermions mix together into ten neutralinos, $\tilde{\chi}_{\alpha}^0$, and five charginos, $\tilde{\chi}_{\alpha}^{\pm}$. Since the three lightest neutralinos are the left-handed neutrinos, the ``true'' LSP would be the $\tilde{\chi}_4^0$. Likewise the three lightest charginos $\tilde\chi^\pm_i$ coincide with the three charged leptons. Similarly, all scalars mix into eight $CP$-even, $S_{\alpha}^0$ and seven $CP$-odd neutral Higgs bosons, $P_{\alpha}^0$,  mass eigenstates. The three lighter neutral scalars, $S_i^0, i=1, 2, 3$, are in fact naturally light singlet-like states, leaving the fourth one, $S_4^0$, to play the role of the discovered SM-Higgs-like scalar. Charged scalars, on the other hand, form seven mass eigenstates, $P_{\alpha}^{\pm}$. Analyses of the \mn, with attention to the neutrino and LHC phenomenology have been addressed in Refs.~\cite{mnssm-spectrum,mnssm-others}. Other analyses concerning cosmology such as gravitino dark matter and electroweak baryogenesis can be found in Refs.~\cite{mnssm-gravitino1,mnssm-gravitino2,mnssm-fermi} and in Ref.~\cite{mnssm-baryo}, respectively. In conclusion, \mn\ is a well-motivated SUSY model with enriched phenomenology and notable signatures, which definitely deserve rigorous analyses by the LHC experiments. Its enlarged Higgs sector can easily accommodate the observed 125~\gev\ Higgs boson~\cite{atlas-higgs,cms-higgs}. 

%%%%%%%%%%%%%%%%%%%%%%%%%%%%%%%%%%%%%%%%%%%%%%%%%%%%%%%%%%%%%%%%%%%%%%%%%%%%%%%%%%%%%%%%%%%%%%%%%%%%
\subsection{Signatures at LHC}\label{sc:munussm-lhc}

Here a collider analysis together with detector simulation of a \mn\ signal featuring non-prompt multileptons at the LHC, arising from the beyond SM decay of a 125~\gev\ scalar into a pair of lightest neutralinos, \XX, is presented~\cite{my-mnssm1}. Since $R$~parity is broken, each \XX\ decays into a scalar/pseudoscalar $(h/P)$ and a neutrino, with the $h/P$ further driven to decay into $\tau^+ \tau^-$, giving rise to a $4\tau$ final state, as shown in Fig.~\ref{fg:decay-chain}. The small $R$-parity breaking coupling of \XX\ renders it long~lived, yet it decays inside the inner tracker, thereby yielding clean detectable signatures: (i) high lepton multiplicity; and (ii) charged tracks originating from displaced vertices (DVs). 

\begin{figure}[htb]
\includegraphics[width=0.32\textwidth]{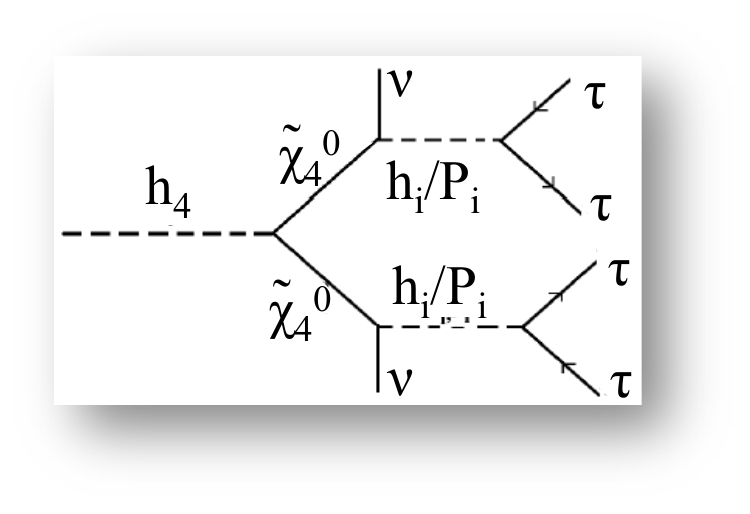}
\hfill
\begin{minipage}[b]{0.63\textwidth}\caption{\label{fg:decay-chain}The \mn\ decay chain studied in Ref.~\cite{my-mnssm1}. The Higgs boson is produced through gluon fusion and has a mass of 125~\gev. The neutralino \XX\ is long-lived and gives rise to displaced $\tau$~leptons.}
\vspace*{0.83cm}
\end{minipage}
\end{figure}

The \mn\ is characterised by the production of several high-\pt\ leptons~\cite{my-mnssm1}. Electrons and muons are produced through the leptonic $\tau$ decays, yet muon pairs can appear directly through $h_i/P_i$ decays as well. In chosen decay mode, the $\tau$ multiplicity is considerable despite the $\tau$-identification efficiency is much lower ($\sim50\%$) when compared to that of electrons and muons ($\gtrsim95\%$). 

Apart from the requirement of at least three or four leptons (including taus), a high value of \met\ and/or of the scalar sum of reconstructed objects: leptons, jets and/or \met\ is needed~\cite{multileptons}. For the chosen signal many neutrinos ($\geq 6$) appearing in the final state from \XX\ and from $\tau$ decay give rise to moderately high ---when compared to signals from RPC SUSY--- \met\ as depicted in Fig.~\ref{fg:met}. Besides \met, the scalar sum of the \pt\ of all reconstructed leptons, $H{\rm _T^\ell}$, can be large in such events. Alternatively, the sum of \met\ and $H{\rm _T^\ell}$ can be deployed for further background rejection. These observables can provide additional handles when selecting events with many leptons. In addition, the invariant masses, $m_{\ell^+\ell^-}$ and $m_{2\ell^+2\ell^-}$ may prove useful for the purpose of signal distinction. 

\begin{figure}[htb]
\begin{minipage}{0.44\textwidth}
\includegraphics[width=\textwidth]{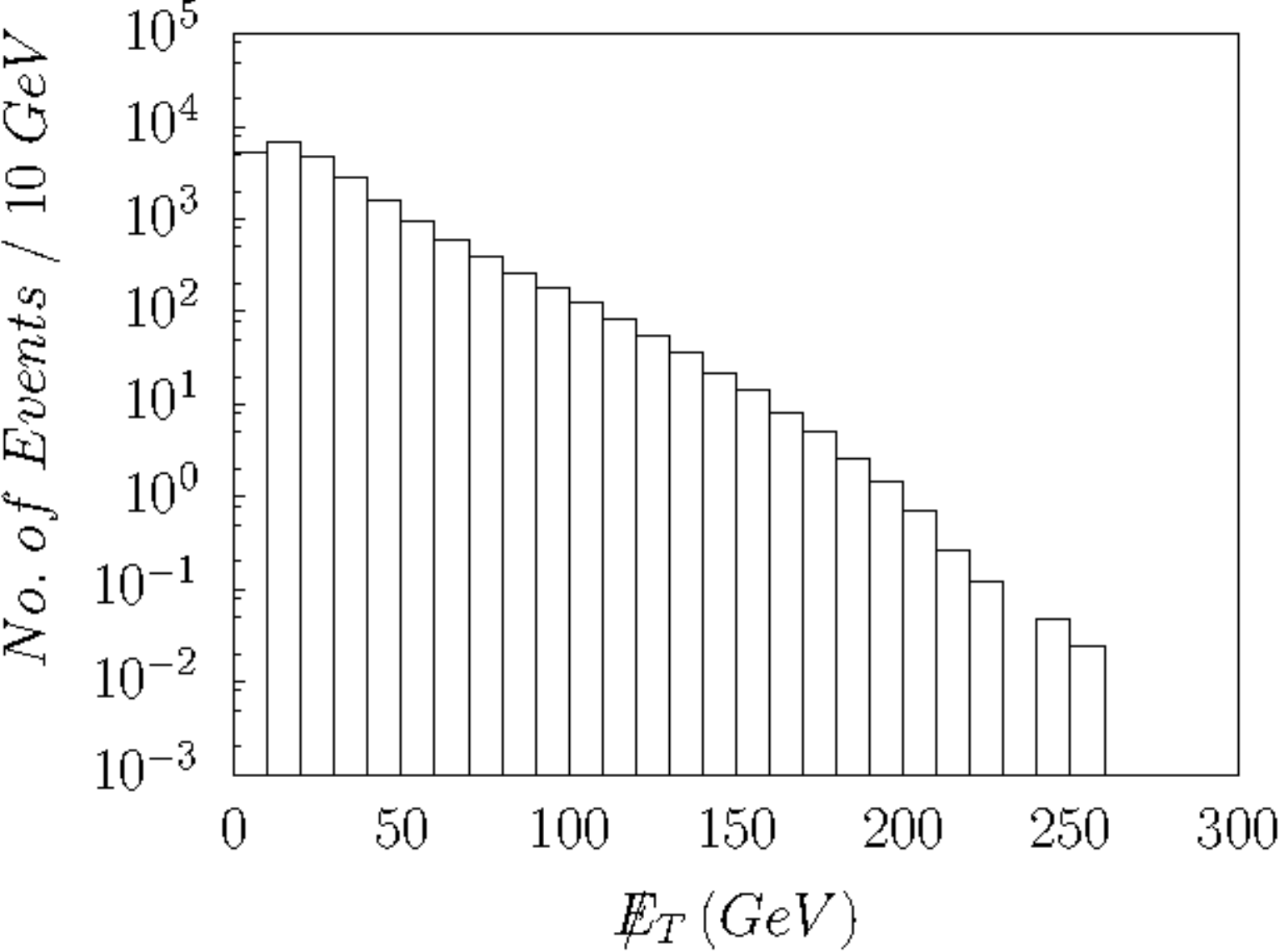}
\caption{\label{fg:met} \met~distribution for an LHC energy of $\sqrt{s}=8~\tev$ and an integrated luminosity $\mathcal{L}=20~\ifb$ for a selected \mn\ point~\cite{my-mnssm1}.}
\end{minipage}
\hfill
\begin{minipage}{0.51\textwidth}
\vspace*{-0.3cm}
\includegraphics[width=\textwidth]{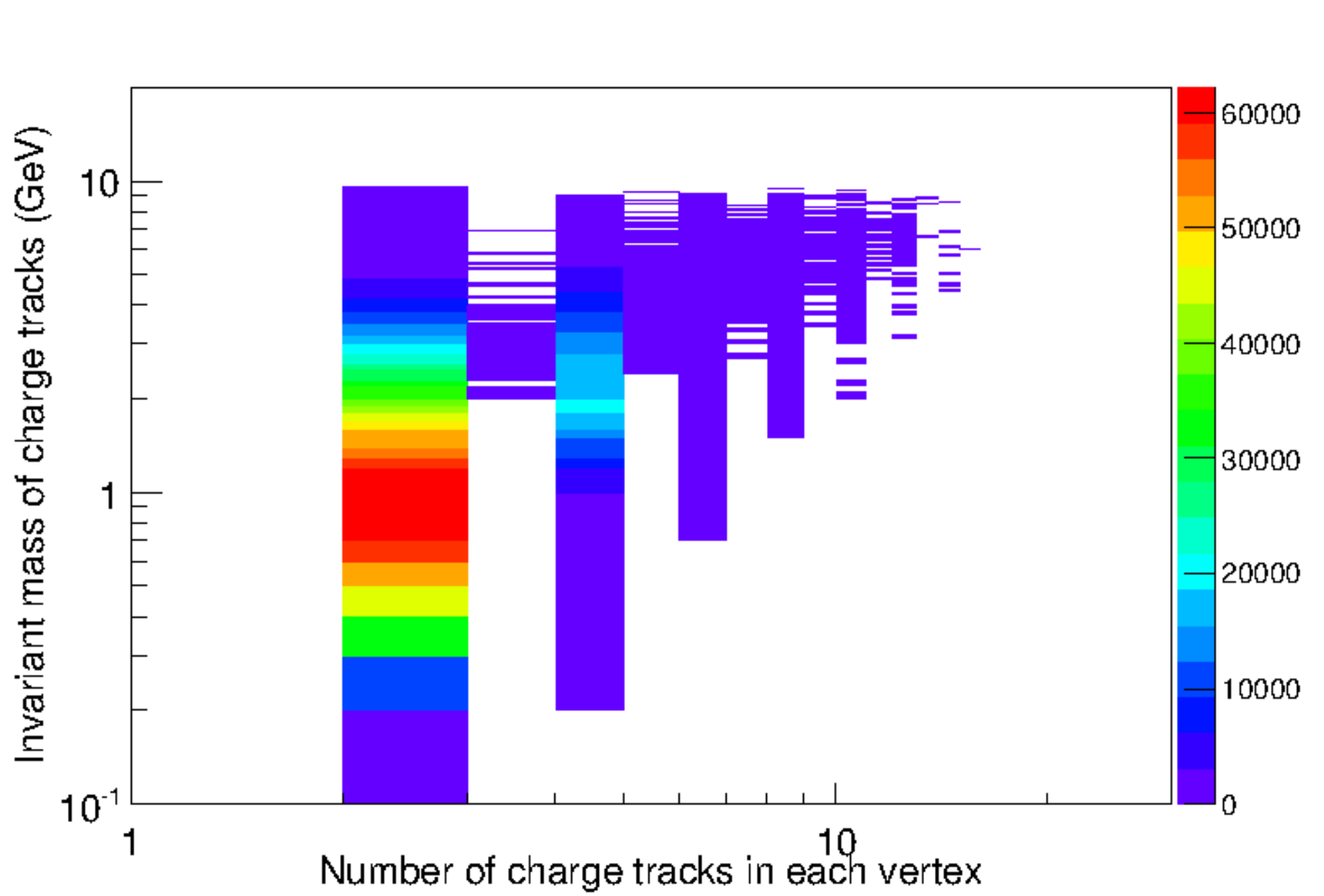}
\caption{\label{fg:DVtrackvsmass12C}Charged-track mass versus the charged-particle number per vertex for LHC at $\sqrt{s}=8~\tev$ and $\mathcal{L}=20~\ifb$ for a selected \mn\ point~\cite{my-mnssm1}.}
\end{minipage} 
\end{figure}

In the benchmark scenario under study, the \XX\ is characterised by a proper lifetime of the order of  $\tau_{\XX}\approx10^{-9}$~s, which corresponds to a proper decay length of $c\tau_{\XX}\approx 30$~cm, thus giving rise to displaced vertices. In a significant percentage of events, the \XX\ would decay inside the inner tracker of the LHC experiments, e.g.\ in 28\% of events it would decay within $30~{\rm cm}$ and in $44$\% events within 1~m. Therefore, the \mn\ signal events would be characterised by displaced $\tau$~leptons plus neutrinos. This distinctive signature opens up the possibility to exploit current or future variations of analyses carried out by ATLAS and CMS looking for a displaced muon and tracks~\cite{atlas-dv,atlas-dv-dilep} or searching for displaced dileptons~\cite{cms-dileptons,atlas-dv-dilep}, dijets~\cite{dijets} or  muon jets~\cite{atlas-dl} arising in Higgs decays to pairs of long-lived invisible particles.  In Fig.~\ref{fg:DVtrackvsmass12C}, the correlation between the number of charged tracks in each DV, $N_{\rm trk}$, and their invariant mass, $m_{\rm DV}$, is shown. The modulation observed in $N_{\rm trk}$ is due to the one-prong or three-prong hadronic $\tau$ decays. 

%%%%%%%%%%%%%%%%%%%%%%%%%%%%%%%%%%%%%%%%%%%%%%%%%%%%%%%%%%%%%%%%%%%%%%%%%%%%%%%%%%%%%%%%%%%%%%%%%%%%
\subsection{Higgs decays}\label{sc:munussm-higgs}

As pointed out earlier, \mn\ can easily predict a Higgs boson of a mass compatible with the observed one~\cite{atlas-higgs,cms-higgs}. As becomes evident from Fig.~\ref{fg:higgs-mass} a 125-\gev\ Higgs boson, $S_4^0$, can be accommodated within a wide range of $\tan\beta$ and $\pmb{\lambda}$ values~\cite{my-mnssm2}. 

\begin{figure}[htb]
\begin{minipage}{0.44\textwidth}
\includegraphics[width=\textwidth]{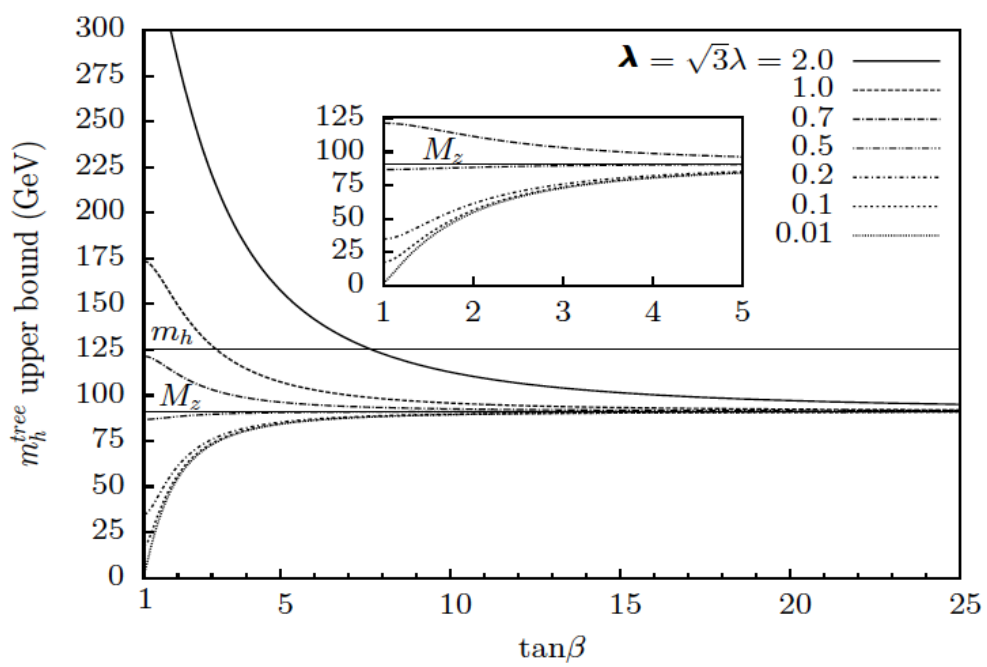}
\end{minipage}
\hfill
\begin{minipage}{0.51\textwidth}
\caption{\label{fg:higgs-mass} Variation of $m_h^{\rm tree}$ upper bound with $\tan\beta$ for different $\pmb{\lambda}$ values in the \mn. The horizontal lines represent the experimentally measured masses of the Higgs, $m_h$, and $Z$~boson, $m_Z$. In the inset the region of $\tan\beta\leq5$ is shown for $\lambda\leq0.7$~\cite{my-mnssm2}.}
\end{minipage} 
\end{figure}

This scalar would produce new states through two-body decays to $S_i^0S_j^0$, $P_i^0P_j^0$, $\tilde{\chi}_{i+3}^0\tilde{\chi}_{j+3}^0$ with $i,j = 1, 2, 3$, as long as these states are kinematically allowed~\cite{my-mnssm2}. It is shown that the final states are dominated by a combination of prompt or displaced leptons, taus, jets and photons plus \met\ due to neutrinos. The expected productions rates for these new signatures are compatible with the measured Higgs signal strengths, $\mu_{XX}$, under the following conditions:
\begin{description}
\item[\underline{$0.01 < \pmb{\lambda} < 0.1$}:] All $\mu_{XX}$ remain within $2\sigma$ of CMS measurements for $2.5 < \tan\beta < 3.9$.
\item[\underline{$0.1 < \pmb{\lambda} < 0.7$}:] Only (invisible) $S_4^0$ decays to $\tilde{\chi}_{i+3}^0\tilde{\chi}_{j+3}^0$ remain viable in the whole range of $\pmb{\lambda}$.
\item[\underline{$\pmb{\lambda} > 0.1$}:] For decays to pair of binos, all $\mu_{XX}$ are within $2\sigma$ for $2.4 < \tan\beta < 3.8$.
\end{description}
Hence there is plenty of room for new Higgs decays within the context of the ``$\mu$ from $\nu$'' supersymmetric standard model. 

%%%%%%%%%%%%%%%%%%%%%%%%%%%%%%%%%%%%%%%%%%%%%%%%%%%%%%%%%%%%%%%%%%%%%%%%%%%%%%%%%%%%%%%%%%%%%%%%%%%%
\subsection{New decays to $Z/W$}\label{sc:munussm-wz}

In addition to the new SM-Higgs-like decays, \mn\ also introduces novel on-shell decays of the $Z$ and $W^{\pm}$ gauge bosons~\cite{my-mnssm3}. These modes are typically encountered in regions of the parameter space populated with light singlet-like scalars, pseudoscalars and neutralinos.  The complete spectrum of possible final states and their origin is presented in Table~\ref{tb:wz}. The delayed ``objects'' occur in delayed decays of the neutralino, whereas the (short-lived) scalars and pseudoscalars deliver prompt products. 

\begin{table}[htb]
\centering
\begin{tabular}{lll}
\hline
 $Z$ decay & & $W^{\pm}$ decay\\
\hline
 $2 x^D 2 {\bar{x}}^D + \met\: (\text{via}~\tilde{\chi}_{i+3}\tilde{\chi}_{j+3})$ &
 & \multirow{2}{*}{$\ell^P/\tau^P + x^D {\bar{x}}^D + \met\: (\text{via}~\tilde\chi^\pm_i\tilde{\chi}_{j+3})$ }\\
 $2 x^P 2 {\bar{x}}^P \: (\text{via}~S^0_iP^0_j)$  & & \\
\hline
\end{tabular}
\caption{Final states from non-standard $Z$ and $W^{\pm}$ decays with their respective origins~\cite{my-mnssm3}. The notation applied is $x: e, \mu, \tau, \gamma, q$ and $P$, $D$ stand for prompt and delayed, respectively.}\label{tb:wz} 
\end{table}

These rare new decays are strongly constrained by the measurements of the $Z$ and $W$ total widths. Signatures  with $\tau$-leptons and/or $b$-jets and with displaced objects would be preferred when probing these decays in order to suppress the huge SM background. Their low production rate, e.g.~$BR\sim\mathcal{O}(10^{-5})$ for the $Z$, necessitates high statistics only becoming available at upcoming collides such as the \emph{GigaZ} and \emph{TeraZ} modes of the Linear Collider and TLEP, respectively~\cite{my-mnssm3}.

%%%%%%%%%%%%%%%%%%%%%%%%%%%%%%%%%%%%%%%%%%%%%%%%%%%%%%%%%%%%%%%%%%%%%%%%%%%%%%%%%%%%%%%%%%%%%%%%%%%%
%%%%%%%%%%%%%%%%%%%%%%%%%%%%%%%%%%%%%%%%%%%%%%%%%%%%%%%%%%%%%%%%%%%%%%%%%%%%%%%%%%%%%%%%%%%%%%%%%%%%
\section{Implications for dark matter}\label{sc:dm}

We address here the issue of (not necessarily cold) dark matter in SUSY models with $R$-parity violation. These seemingly incompatible concepts \emph{can} be reconciled in bRPV models with a gravitino~\cite{gravitino,martin,brpv-dm,brpv-split,ams02-brpv} or an axino~\cite{axino} LSP with a lifetime exceeding the age of the Universe. In both cases, RPV is induced by bilinear terms in the superpotential that can also explain current measurements on neutrino masses and mixings without invoking any GUT-scale physics. Decays of the next-to-lightest superparticle occur rapidly via RPV interactions, thus they do not disturb the Big-Bang nucleosynthesis, unlike the $R$-parity conserving case~\cite{leptog}. Decays of the NLSP into the gravitino and SM particles do not contribute to the gravitino relic density in scenarios with broken $R$~parity. This is due to decay processes involving a gravitino in the final state that are suppressed compared to $R$-parity breaking decays unless the amount of $R$-parity violation is extremely small.

Evidence on the four-year Fermi data that have found excess of a 130~\gev\ gamma-ray line from the Galactic Centre (GC) have been studied in the framework of $R$-parity breaking SUSY. A decaying axino DM scenario based on the SUSY KSVZ axion model with the bilinear $R$-parity violation explains the Fermi 130-\gev\ gamma-ray line excess from the GC while satisfying other cosmological constraints~\cite{brpv-axino}. On the other hand, gravitino dark matter with trilinear RPV can account for the gamma-ray line, since there is no overproduction of anti-proton flux, while being consistent with Big-Bang nucleosynthesis and thermal leptogenesis~\cite{trpv-gravitino}.

Measurements of the cosmic-ray antiproton flux by PAMELA, Fermi-LAT and AMS-02 have been used recently to constrain the (decaying) gravitino mass and lifetime in the channels $Z\nu$, $W\ell$ and $h\nu$~\cite{ams02-brpv,grefe}. Subsequently upper limits have been set on the size of the $R$-parity violating coupling in the range of $10^8$ to $8\times10^{13}$. 

Recent analyses of multiple galaxy-cluster spectra and the Andromeda galaxy from the XMM-Newton telescope, have revealed a tentative line with the central energy of 3.5~keV. Any long lived particle that produces enough number of photons should be a good candidate as a source of X-rays. Among the various theoretical scenarios attempting to explain it, RPV SUSY with an LSP decaying to photon and a neutrino is a possibility. In bilinear RPV it was found that a warm-dark-matter axino  with a mass of $m_{\tilde{\alpha}}\simeq7~{\rm keV}$ can have the proper lifetime and relic density to account for the observed X-ray emission line through its decay $\tilde{\alpha}\to\gamma\nu$~\cite{axino-brpv,xray35-brpv}. The axino parameter space consistent with the aforesaid X-ray line can be seen in Fig.~\ref{fg:axino-brpv}. Alternatively the line may be due to annihilating DM with $m_{\tilde{\alpha}}\simeq3.5~{\rm keV}$. Apart from the axino, other possible candidates include a gravitino, a bino, or a hidden sector photino as decaying DM leading to such a signal studied in the context of other-than-bilinear RPV SUSY scenarios~\cite{xray35-other}.

\begin{figure}[htb]
\centering
\vspace*{-0.3cm}
\includegraphics[width=0.6\textwidth,clip=]{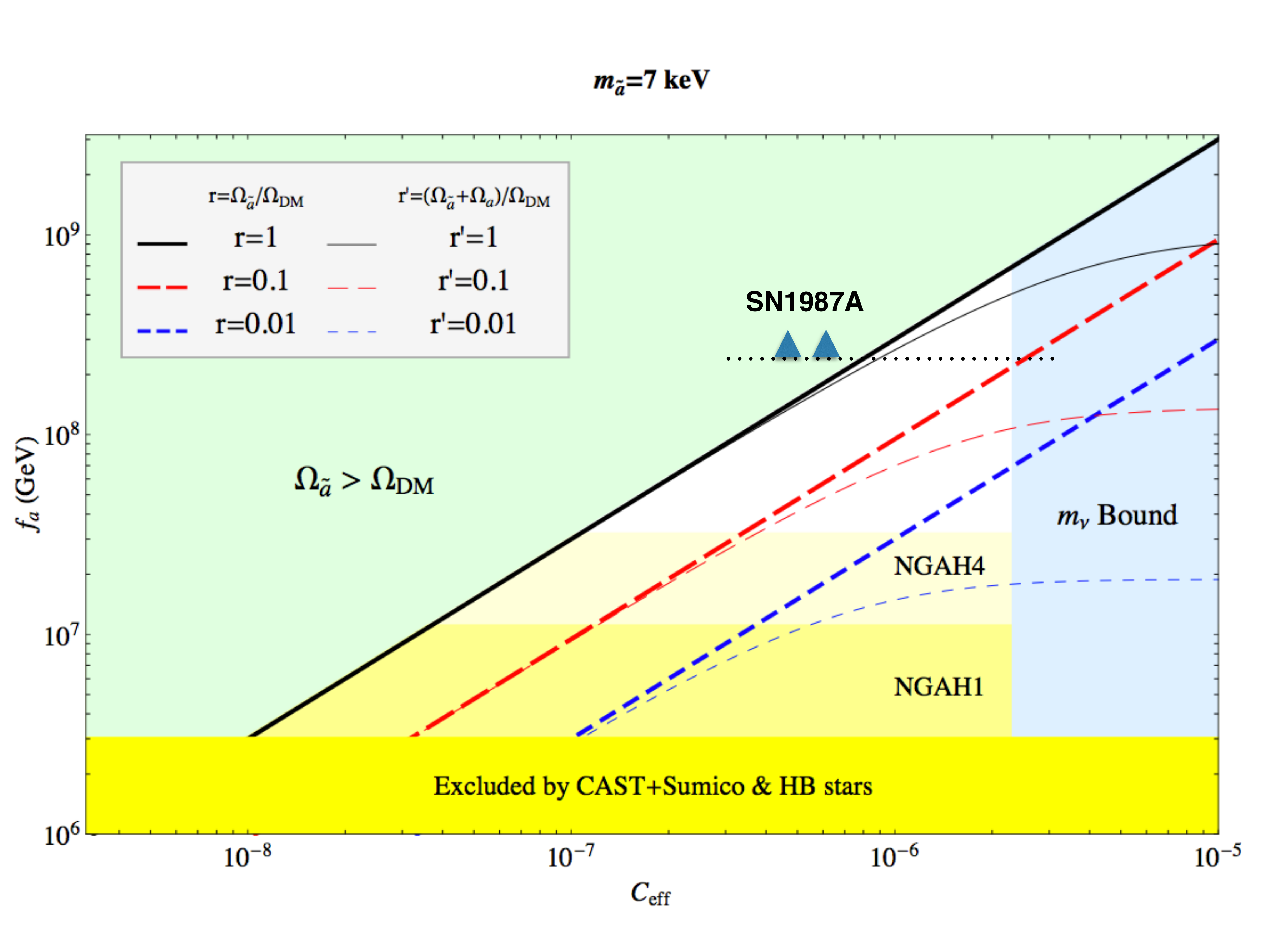}
\caption{\label{fg:axino-brpv} Parameter space consistent with the 3.5~keV line signal in the $(C_{\rm eff},\,f_a)$ plane.
The thick (thin) lines represent the values of $C_{\rm eff}$ and $f_a$ to fit the required lifetime for different values of $r$ ($r'$).
The upper (green), lower (yellow), and right (blue) shaded regions are excluded by the DM relic density, axion-like particle searches and astrophysical observations, and the neutrino mass limit, respectively.
Conservative projected limits from NGAH are shown as the (light-yellow) shaded regions. From Ref.~\cite{axino-brpv}. }
\end{figure}

Such gravitino DM is also proposed in the context of \mn\ with profound prospects for detecting $\gamma$ rays from their decay~\cite{mnssm-gravitino1}. In studies on the prospects of the Fermi-LAT telescope to detect such monochromatic lines in observations of the most massive nearby extragalactic objects, it was found that a gravitino with a mass range of $0.6-2~\gev$, and with a lifetime range of about $3\times10^{27} - 2\times10^{28}$~s should be detectable with a signal-to-noise ratio larger than three~\cite{mnssm-gravitino2}. After confronting the actual $\gamma$-ray flux data, limits on the model parameters have been set~\cite{mnssm-gravitino2,mnssm-fermi}. 

$R$-parity breaking couplings can be sufficiently large to lead to interesting expectations in colliders. The neutralino NLSP, depending on the RPV model, may decay into~\cite{martin,ll-higgsinos}
\begin{eqnarray} \nonumber
& \X\to W^{\pm}\ell^{\mp},       & \qquad \X\to\t{G}\gamma, \\ 
& \X\to \nu\tau^{\pm}\ell^{\mp}, & \qquad \X\to\t{G}Z,      \\ \nonumber
& \X\to \nu\gamma,               &  \qquad \X\to\t{G}h.
\end{eqnarray}

Such decays may be probed at the LHC via inclusive channels characterised by leptons, many jets, large \met\ and/or photons or exclusive channels involving the reconstruction of a $Z$ or a $h$ from its decay products. The cases where the NLSP is long-lived yet with a decay length comparable to the dimensions of an LHC experiment are particularly interesting, as they give rise to displaced tracks/leptons and non-pointing photons. The possibility to measure the neutralino decay length provides an extra handle to constrain the underlying SUSY model. 

%%%%%%%%%%%%%%%%%%%%%%%%%%%%%%%%%%%%%%%%%%%%%%%%%%%%%%%%%%%%%%%%%%%%%%%%%%%%%%%%%%%%%%%%%%%%%%%%%%%%
%%%%%%%%%%%%%%%%%%%%%%%%%%%%%%%%%%%%%%%%%%%%%%%%%%%%%%%%%%%%%%%%%%%%%%%%%%%%%%%%%%%%%%%%%%%%%%%%%%%%
\section{Summary and outlook}\label{sc:summary}

The hitherto null results of searches for supersymmetry in conventional channels calls for a more systematic and thorough consideration of non-standard SUSY theoretical scenarios and experimental techniques. To this effect, scenarios involving violations of $R$~parity and/or (meta)stable particles arise as interesting alternatives.

$R$-parity violating supersymmetry can reproduce correctly the measured neutrino physics parameters. Moreover its enriched mass spectrum and $R$-parity breaking decays can lead to novel signals at colliders, among which very few possibilities have been explored so far. In particular, searches for displaced objects at the LHC offer an attractive possibility as well as the prompt multilepton analyses. The former suffer from less background sources from SM processes when compared to prompt-object searches. More sophisticated searches are expected by ATLAS, CMS and LHCb soon as the first data from LHC Run~II at $13~\tev$ are being analysed. 

%%%%%%%%%%%%%%%%%%%%%%%%%%%%%%%%%%%%%%%%%%%%%%%%%%%%%%%%%%%%%%%%%%%%%%%%%%%%%%%%%%%%%%%%%%%%%%%%%%%%
%%%%%%%%%%%%%%%%%%%%%%%%%%%%%%%%%%%%%%%%%%%%%%%%%%%%%%%%%%%%%%%%%%%%%%%%%%%%%%%%%%%%%%%%%%%%%%%%%%%%
\section*{Acknowledgements}

The author is grateful to the PLANCK~2015 organisers for the invitation to the Conference. She thanks Pradipta Ghosh for providing material for the paper. She acknowledges support by the Spanish Ministry of Economy and Competitiveness (MINECO) under the project FPA2012-39055-C02-01, by the Generalitat Valenciana through the project PROMETEO~II/2013-017 and by the Spanish National Research Council (CSIC) under the JAE-Doc program co-funded by the European Social Fund (ESF). The author benefited from the CERN Corresponding Associate Programme.

%%%%%%%%%%%%%%%%%%%%%%%%%%%%%%%%%%%%%%%%%%%%%%%%%%%%%%%%%%%%%%%%%%%%%%%%%%%%%%%%%%%%%%%%%%%%%%%%%%%%
%%%%%%%%%%%%%%%%%%%%%%%%%%%%%%%%%%%%%%%%%%%%%%%%%%%%%%%%%%%%%%%%%%%%%%%%%%%%%%%%%%%%%%%%%%%%%%%%%%%%

\end{document}